\documentclass{article}
\usepackage{spconf,amsmath,graphicx}
\usepackage{seqsplit}
\usepackage{caption}
\usepackage[inline]{enumitem} 
\usepackage{mathtools}
\usepackage{amsfonts}
\usepackage{algorithm}
\usepackage{algorithmic}
\usepackage{booktabs}
\usepackage{subcaption}
\usepackage{hyperref}
\usepackage{tabularx}
\usepackage{booktabs}
\usepackage{multirow}
\newcommand{\T}{\mathcal{T}}
\newcommand{\xs}{x_\textup{source}}
\newcommand{\xsh}{\hat{x}_\textup{source}}
\newcommand{\xt}{x_\textup{target}}
\newcommand{\STIMIT}{\mathcal{S}_{\textup{TIMIT}}}

\setlist[itemize]{topsep=1pt}
\title{MicAugment: One-shot Microphone Style Transfer}
\name{Zal\'an Borsos$^{\dagger,1}$\thanks{$^1$Work done while interning at Google.} \qquad Yunpeng Li$^{\star}$  \qquad Beat Gfeller$^{\star}$ \qquad  Marco Tagliasacchi$^{\star}$}
\address{$^{\dagger}$ Department of Computer Science, ETH Zurich \\
      $^{\star}$ Google Research \\
      zborsos@ethz.ch \quad \{yunpeng, beatg, mtagliasacchi\}@google.com}

\begin{document}

\setlength{\abovedisplayskip}{4.6pt}
\setlength{\belowdisplayskip}{4.6pt}

%\ninept
%
\maketitle
\begin{abstract}
A crucial aspect for the successful deployment of audio-based models ``in-the-wild'' is the robustness to the transformations introduced by heterogeneous acquisition conditions. In this work, we propose a method to perform \emph{one-shot microphone style} transfer. Given only a \emph{few seconds} of audio recorded by a target device, \emph{MicAugment} identifies the transformations associated to the input acquisition pipeline and uses the learned  transformations to synthesize audio as if it were recorded under the same conditions as the target audio. We show that our method can successfully apply the style transfer to real audio and that it significantly increases model robustness when used as \emph{data augmentation} in the downstream tasks.
\end{abstract}
\begin{keywords}
style transfer, robustness, data augmentation
\end{keywords}

\section{Introduction}

\looseness -1 Recent advances in audio recognition models produced significant performance improvements in a number of tasks.
Yet these systems can witness severe performance degradation when encountering \emph{domain shift}~\cite{8732945}. This is often the case in practice: the deployed  models face audio data collected in various environments by heterogeneous devices, each characterized by different hardware and signal processing pipelines. 

\looseness -1 A prominent approach for tackling domain shift is the development of robust models. In the speech recognition community, the topic of model robustness has a long history~\cite{junqua2012robustness}.
Closely related to our work is robustness to microphone variability,  reverberation and  noise. 
Approaches for increasing robustness to reverberation and noise fall into two main categories: methods that focus on dereverberation and denoising \cite{4749462} and methods that enforce robustness by training with data collected in diverse environmental conditions  \cite{6747994}. 
Concerning the robustness to microphone variability,  pioneering methods using additive corrections in the cepstral domain~\cite{acero1990environmental} and multi-style training using a collection of microphones~\cite{acero2012acoustical} were proposed.
Recently, \cite{mathur2018using} proposed a data augmentation method based on relative microphone transfer functions learned from paired and aligned data from multiple devices. However, the assumption on the availability of paired data hinders the practicality of the approach. In~\cite{8732945}, the authors propose to learn mappings  between microphones using CycleGAN~\cite{zhu2017unpaired}. 
While working in the unpaired setup without assumptions on the microphone model, the method relies on training a separate CycleGAN for every microphone type encountered during test time. The method also inherits the weaknesses of CycleGAN, e.g., mapping  unrelated samples in the source and target domains.

\looseness -1 In this work, we investigate the following problem: given a \emph{few seconds} of audio recorded in the target environment, transform other audio samples to sound as if they were recorded in the target environment. We focus on environmental transformations due to \emph{microphone variability}, early room \emph{reverberation} and \emph{noise}.
To this end, focusing on \emph{speech} data, we propose \emph{MicAgument}, a novel approach to one-shot microphone style transfer. 
Operating in the time-domain, the method is agnostic to the downstream tasks, and hence widely applicable. Our main contributions are the following:
\begin{itemize}\setlength\itemsep{-0.3em}
    \item \looseness -1 We develop a lightweight non-linear microphone model, the core component of MicAugment, based on strong inductive biases capturing the microphone signal processing pipeline, that can be inferred using just a few seconds of audio from the target device;
    \item We show that the samples transformed by MicAugment can fool a device identification model; 
    \item Finally, we demonstrate that, when used as data augmentation, MicAugment can significantly increase model robustness to microphone variability.
\end{itemize}

\section{Method}\label{sec:method}
\looseness -1 For the development of our proposed method, we rely on the availability of the following data: \begin{enumerate*}[label=\roman*)]
  \item  a collection of speech samples, which we refer to as \emph{source},
  \item a short speech segment recorded by the device of interest, which we refer to as \emph{target}.
\end{enumerate*} We assume that the source samples contain \emph{clean} speech samples, collected with a high-quality microphone with flat frequency response and only mildly affected by background noise and room reverberation. This assumption is satisfied by commonly used corpora such as VCTK~\cite{yamagishi2019vctk} or Librispeech~\cite{panayotov2015librispeech}. 
In this setup, the output of MicAugment is a transformation that modifies the source samples to sound as if they were recorded by the target microphone.

\begin{figure}[t!]
\centering
\includegraphics[width=0.35\textwidth]{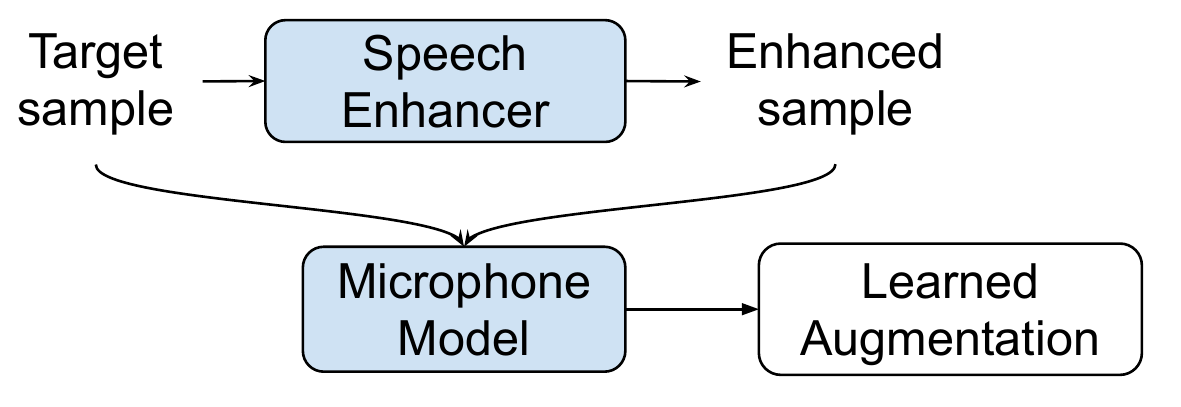}
\vspace{-2mm}
\caption{Overview of MicAugment.}
\vspace{-5mm}
\label{fig:overview}
\end{figure}

 The overview of MicAugment is illustrated in Figure~\ref{fig:overview}. The first building block is the \emph{speech enhancement network} recently proposed by \cite{tagliasacchi2020seanet}.
We find that the speech enhancement network is successful in extending frequencies lost due to  microphone filtering, in removing noise and even early room reverberation. However, robustness to microphone transformations can not be achieved with the speech enhancement network alone: when deployed in diverse environments, its  output is not artifact-free --- we observe on real datasets that the introduced artifacts significantly degrade the performance of models in the downstream tasks. Moreover, running the enhancement network at inference time would introduce additional latency.

Therefore, inspired by the recent success of relying on inductive biases in the audio domain~\cite{Engel2020DDSP}, we propose a \emph{microphone model} 
by leveraging strong inductive biases in the signal acquisition pipeline. 
These strong priors are crucial for the success of the method: they allow for model identification given only a few seconds of audio and even in the presence of artifacts possibly introduced by the speech enhancement network. The identified transformations can then be used for data augmentation during training to achieve robustness to the target domains. In the following, we present the individual components of our approach.

\subsection{Microphone Model}

The task of the microphone model is to approximate the transformation $\T$ attributed to the input acquisition conditions. We propose to approximate $\T$ based on a single pair $(\xs, \xt)$ containing just a few seconds of speech, under the assumption that $\xt = \T(\xs)$. To achieve this, we incorporate strong inductive biases in modeling the audio acquisition pipeline, yet flexible enough to obtain convincing experimental results.

\begin{figure}[t!]
\centering
\includegraphics[width=0.48\textwidth]{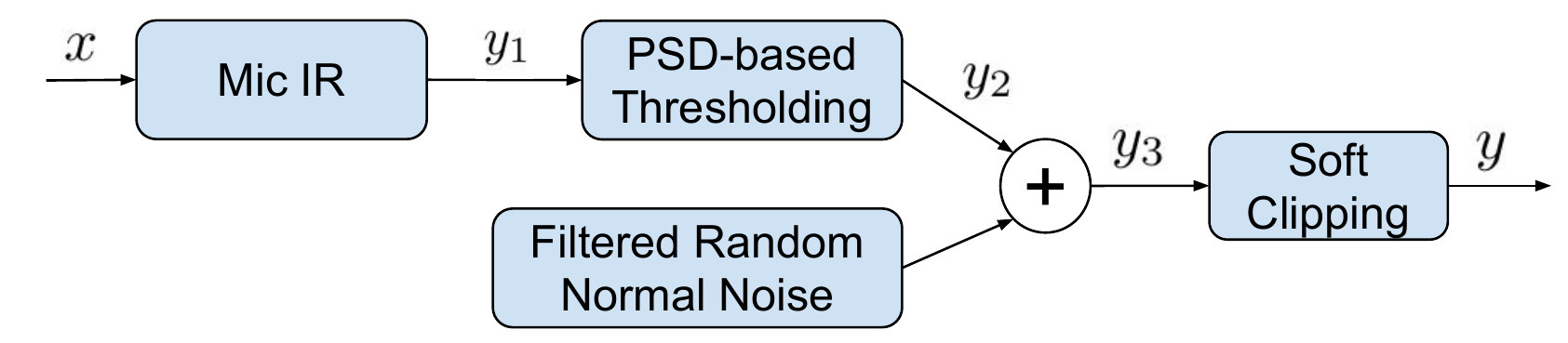}
\vspace{-6mm}
\caption{The microphone model and its components.}
\label{fig:mic-model}
\vspace{-5mm}
\end{figure}

The microphone model is shown in Figure~\ref{fig:mic-model}. As the first step of the pipeline, the input time-domain waveform $x\in \mathbb{R}^N$ is convolved with the microphone impulse response (IR) $f_m$:
\begin{equation}
y_1  = f_m * x.
\end{equation}
In our experiments, we found that \emph{robustness to room reverberation} is crucial in several tasks. Hence, with a slight abuse of nomenclature, we denote with $f_m$ the composition of the microphone IR and the room IR.
In order to handle variable-sized inputs of the order of thousands of samples, we perform all  convolutions as multiplications in the frequency domain with time complexity $\mathcal{O}(N \log N)$.

\looseness -1 In the second step, $y_1$ is passed through a component that models the non-linearity of the device audio processing pipeline by frequency band cutouts based on the power spectral density (PSD): when the signal power in a frequency band is lower than a learned threshold, the respective power is further decreased in that band. Formally,
\begin{equation} \label{eq:psd-thresholding}
y_2  = \textup{STFT}^{-1} \left(\textup{STFT}(y_1) \cdot \sigma(|\textup{STFT}(y_1)|^2 - t ) \right),
\end{equation}
where $\sigma(\cdot)$ is the sigmoid function and $t$ is broadcast over time, having dimension corresponding to the STFT window length. Next, we sample a white Gaussian noise that is convolved with a learned filter $f_n$ and added to $y_2$ to produce $y_3$,
\begin{equation}
y_3 = y_2 + f_n * \epsilon, \quad \epsilon \sim \mathcal{N}(0,1).
\end{equation}
Our last component models the microphone clipping effect by a learneable soft-threshold for the maximum absolute value for the waveform in a differentiable manner:
\begin{equation} \label{eq:clipping}
 y = \textup{smoothmin}(\textup{smoothmax}(y_3, -\tau), \tau),
\end{equation}
where $\textup{smoothmax}(a, b) = (ae^a +be^b) / (e^a +e^b)$ and $\textup{smoothmin}(a, b) = (ae^{-a} +be^{-b}) / (e^{-a} +e^{-b})$ are applied elementwise. Let us denote the set of learnable parameters by $\theta =\{f_m, f_n, t, \tau \}$ and the resulting microphone transformation by $\textup{mic}_\theta(\cdot)$. We propose to optimize the microphone model's parameters $\theta$ given just a few seconds of paired audio samples $(\xs, \xt)$ by  gradient descent on the objective
\begin{equation}
    \min_\theta \ell_1 \left(\textup{MEL}(\textup{mic}_\theta(\xs)),\,  \textup{MEL}(\xt) \right),
\end{equation}
where $\ell_1(\cdot, \cdot)$ denotes the mean absolute error and $\textup{MEL}(\cdot)$ computes the log-mel spectrogram.

\subsection{Speech Enhancement Network}

\looseness -1 The estimation of the microphone model assumes that the input audio $\xs$ represents clean speech, which is often unavailable in practice. We therefore resort to a speech enhancement network to produce an approximation $\xsh = E(\xt)$, by enhancing the sample $\xt$  through inverting the microphone transformations encountered in the target domain. Since $\xt$ is possibly affected by noise, microphone filtering and room reverberation, the task of this network is to perform both blind deconvolution and denoising.

\looseness -1 We create a synthetic training dataset for the enhancement network that covers a wide variety of transformations:  the clean reference audio is the VCTK corpus~\cite{yamagishi2019vctk}, containing sentences from 56 native English speakers, split into 1-second segments; each segment is modified by applying transformations randomly sampled from our microphone model. This is possible due to the microphone model's  modular construction presented in Figure~\ref{fig:mic-model}, where the components can be sampled individually. 

For sampling the microphone model, we collect a set of 70 microphone IRs from the MICrophone Impulse Response Project~\cite{micirp}, complemented with a set of 200 band-pass filters with cutoff frequencies uniformly sampled from [50 Hz, 150 Hz] and [3 kHz, 8 kHz]. 
We collect a set of 1000 room reverberation IRs by running a room simulator based on the image-source method on rooms of diverse geometry. Finally, we extract $10000$ noise segments from the Speech Commands dataset~\cite{speechcommandsv2}, by applying the same extraction method as in~\cite{li2020learning}, i.e., by searching for short low-energy segments of 100 ms length and replicating them (randomizing the phase) with overlap-and-add to obtain samples of 1 second.

We create a parallel corpus by applying the following operations to each VCTK sample: 
\begin{enumerate*}[label=\roman*)]
  \item convolution with a randomly chosen room reverb IR;
  \item convolution with a randomly chosen microphone IR;
  \item thresholding using Eq.~\eqref{eq:psd-thresholding}, by dividing the frequency range into 8 equal buckets and sampling the thresholds uniformly at random from the range $[-2, 3]$  for each bucket;
  \item adding a randomly chosen noise sample with adjusted gain such that the resulting SNR is in the range [5 dB, 30 dB]; and, finally,
  \item clipping using Eq.~\eqref{eq:clipping} where $\tau$ is chosen randomly between the half and the maximum absolute time-domain signal value.
\end{enumerate*}
Each of the previous operations is applied with a probability of 0.8, 0.9, 0.6, 0.9 and 0.1, respectively, where the values have been chosen to ensure diversity in the resulting samples. 

The architecture of the enhancement network consists of a fully-convolutional wave-to-wave U-Net identical to the audio-only version of SEANet~\cite{tagliasacchi2020seanet}. We follow the same training setup, by minimizing a loss function which is a combination of an adversarial loss and a feature matching loss, where features are extracted from the intermediate activations of the multi-scale discriminator proposed in MelGAN~\cite{kumar2019melgan}.

\section{Experiments}\label{sec:experiments}
 In this work, we propose to asses the quality of the style-transferred samples via a series of \emph{downstream tasks}, i.e., supervised learning problems that receive the transferred samples as inputs. In the experiments, we employ two downstream tasks: i) fooling a device identification model and ii) evaluating the robustness of a fully supervised model trained with different augmentation strategies.  While this evaluation method is inherently dependent on the models used in the downstream tasks, we find that the relative ordering of the competing methods is preserved across different tasks.
 
 \looseness -1 In the experiments, all signals are assumed to be sampled at 16 kHz. The mel spectrogram is calculated with a window length of 1024 samples (64 ms) and a hop length of 160 (10 ms) and with 128 bins. The optimization of the microphone model is performed with ADAM~\cite{kingma2015adam} using step size 0.005 and 1000 iterations, which takes less than a minute on a single GPU due to the small number of parameters. We fix the STFT window length to 2048 (128 ms) and hop length to 160. Both the speech enhancement network and the microphone model operate on gain normalized signals.

\subsection{Device Identification} \label{subsec:attacking-device-id}
\looseness -1 We consider the problem of identifying a mobile device from the footprints left by its microphone,  a topic of interest in audio forensics~\cite{kraetzer2007digital}. In this experiment, we start by training a fully supervised device identification model. Then, given only a few seconds of audio from a target device, we apply MicAugment to the clean samples with the goal of fooling the device identification model to believe that they were taken from the target device.  

For this task, we use the MOBIPHONE dataset~\cite{6900732} recording sentences of 12 male and 12 female speakers randomly chosen from TIMIT database~\cite{garofolo1993darpa} with 21 mobile devices, resulting in $~30$ seconds of audio per speaker and device. Following the setup in~\cite{verma2019cnn},  we exclude the phone ``Samsung s5830i'' due to the small sample size, resulting in a 20-class classification task. We split the dataset into train and test sets with non-overlapping sets of 16 and 8 speakers and sentences. The audio is processed in chunks of one second.
We also create a paired and aligned TIMIT-MOBIPHONE dataset, which allows us to evaluate our microphone model in isolation.

\begin{table*}[t]
\begin{minipage}{0.28\linewidth}
\centering
\caption{Success rates of style transfer in fooling the device identifier. Mean and std. dev. over 5 random $\STIMIT$ batches and  seeds.\label{table:mobiphone-attack}}
% \vspace{-1mm}
\begin{tabular}{@{}cc@{}}
\toprule
\textbf{Method}                          & \textbf{Success $\%$} \\ \midrule
\textbf{Spectr. Eq.}               & 34.4$\pm$1.3               \\
\textbf{Mic. Model (paired)} & 88.3$\pm$0.6               \\
\textbf{MicAugment}           & 68.3$\pm$2.4                \\ \bottomrule
\end{tabular}
\vspace{-2mm}
\end{minipage}
\hfill 
\begin{minipage}{0.42\linewidth}
\centering
\caption{Robustness to microphone variability.\label{table:speech-comm}}
\vspace{-2mm}
\begin{tabular}{ccc}
\toprule
\multirow{2}{*}{\textbf{\begin{tabular}[c]{@{}c@{}} \\ Method \end{tabular}}} & \multicolumn{2}{c}{\textbf{Test Acc. Speech Commands~\cite{speechcommandsv2}}}                                                                                                   \\ \cline{2-3} 
                                 & \textbf{\begin{tabular}[c]{@{}c@{}}Synth. / \\ Recovery \%\end{tabular}} & \textbf{\begin{tabular}[c]{@{}c@{}}Real /\\ Recovery \%\end{tabular}} \\ \midrule
\textbf{No Augm.}                & 91.8$\pm$0.3 / 0                                                        & 88.3$\pm$0.3 / 0                                                       \\
\textbf{Spectr. Eq.}            & 92.5$\pm$0.1 / 25.0                                                     & 89.3$\pm$1.6 / 35.6                                                    \\
\textbf{SpecAugm.}               & 92.4$\pm$0.2 / 20.5                                                     & 89.0$\pm$0.2 / 26.1                                                    \\
\textbf{MicAugm.}                & 93.8$\pm$0.1 / 71.6                                                     & 90.3$\pm$0.2 / 73.0                                                    \\
\textbf{Oracle}                  & 94.5$\pm$0.1 / 100                                                      & 91.0$\pm$0.3 / 100                                                       \\ \bottomrule
\end{tabular}
\vspace{-3mm}
\end{minipage}
\hfill 
\begin{minipage}{0.28\linewidth}
\centering
\includegraphics[width=0.98\textwidth]{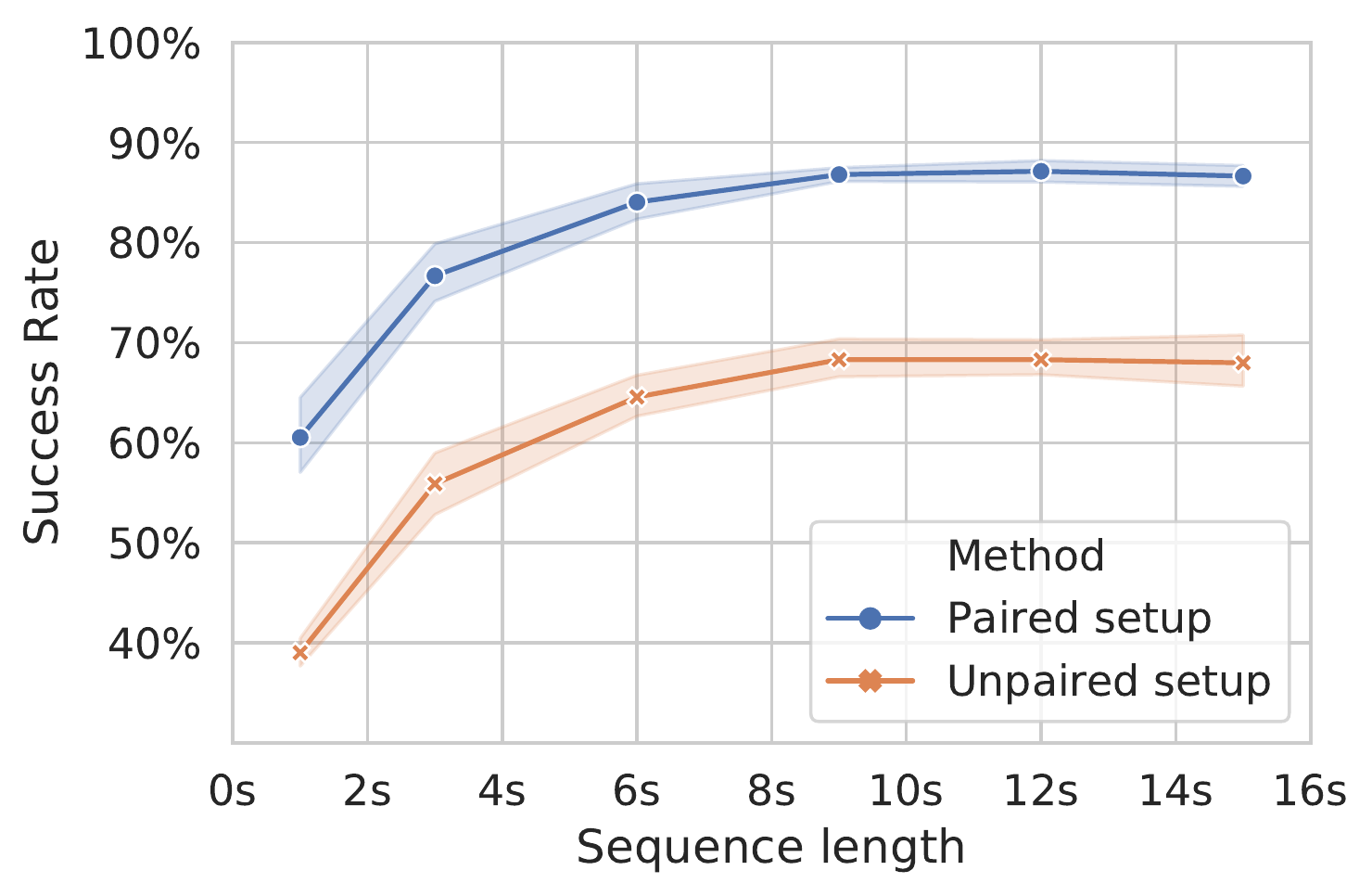}
\vspace{-3mm}
\captionof{figure}{The effect of target audio length on the success rate of fooling the device identification model. }
\label{fig:ablation-length}
\end{minipage}
\vspace{-3mm}
\end{table*}

\looseness -1 Our device identification model is a convolutional neural network (CNN) operating on log-mel spectrogram inputs (window length 25 ms, hop length 10 ms, 64 mel-channels). The CNN is composed of six convolutional blocks, each consisting of separable convolutions along the time and the frequency axes with $3\times3$ kernels followed by ReLU and interleaved with batch normalization. The number of channels in each block is equal to [64, 128, 256, 256, 512, 512]. Max pooling with a stride of 2 along both time and frequency is applied when the number of channels is increased. The convolutional body is followed by a head with two layers: a fully connected layer with 256 outputs with ReLU, and a final linear layer with 20 outputs. 
This model achieves $99.2\%$ test accuracy averaged over the  one-second chunks, and $100\%$ test accuracy when majority voting is applied for each speaker and device, a slight improvement over the results of~\cite{verma2019cnn}. This model serves as the basis for the attacks via style transfer.

For constructing our baselines, we sample $\STIMIT$ by choosing 100 random  speakers from TIMIT and retrieving a random sentence with an average length of 3 seconds for each speaker. The goal of the methods is to propose transformations \emph{for each target device}, such that when applied to $\STIMIT$, the  transformed samples are misclassified by the CNN as belonging to the target device. We report the success rate averaged over both the  devices and the samples in $\STIMIT$. 

\looseness -1 We propose two baselines that rely on the paired and aligned TIMIT-MOBIPHONE datasets. For training,  we select 15 seconds of paired audio for each device, containing one sentence from 5 speakers chosen at random. The first baseline is spectral equalization~\cite{stockham1975blind}, i.e., calibration  based on the power spectral density (PSD) ratios estimated from the TIMIT and MOBIPHONE batches: the method adjusts the PSD of $\STIMIT$ to match the PSD of the samples from the target device, hence capturing only linear transformations.  The second baseline is learning our proposed microphone model from the 15 seconds of paired data for each device. Finally, we showcase MicAugment, which works in the unpaired setup and has access only to the target device recordings.

\looseness -1 The results in Table \ref{table:mobiphone-attack} show that in the paired setup, the microphone model learns transformations that fool the device identifier with $88.3\%$ success rate. MicAugment succeeds $68.3\%$ of the time when provided with only 15 seconds of audio samples from each target device. We perform an ablation study on the effect of the length of audio data from target device on the success rate of fooling the device identification model. The results are shown in Figure~\ref{fig:ablation-length}, where we observe that the peak performance is already reached with 9 seconds of audio. The results indicate that the $20\%$ of accuracy drop compared to the paired setup is due to the speech enhancement network, which faces real data transformation not encountered during training time. A collection of source and  target samples, as well as samples transformed with MicAugment are provided in the accompanying material.\footnote{\footnotesize \url{https://google-research.github.io/seanet/micaugment/examples}}

\subsection{Keyword Detection}
\looseness -1 In this experiment, we illustrate the accuracy degradation due to microphone variability in a keyword detection systems and show how it can be mitigated by data augmentation with  MicAugment. We first train the same CNN as in Section~\ref{subsec:attacking-device-id} for the keyword detection task on the Speech Commands V2 dataset~\cite{speechcommandsv2}. We then measure the accuracy drop of the classifier when the test samples are captured via different devices. To this end, we consider two scenarios: 
\begin{enumerate*}[label=\roman*)]
  \item \emph{synthetic setup}, where the test set is transformed via the microphone models learned in the previous section on the TIMIT-MOBIPHONE paired setup,
  \item \emph{real setup}, where the test set is recaptured playing the audio clips with MacBook Pro loudspeakers and recording them with the built-in microphone.
\end{enumerate*}
The first row of Table~\ref{table:speech-comm} shows the accuracy drop from 95.7\% to either 91.8\% or 88.3\% depending on the test scenario.

To obtain an upper bound on the accuracy achievable during test, we learn an oracle model which jointly leverages during training the same augmentations applied at test time, namely:  
\begin{enumerate*}[label=\roman*)]
  \item all transformations in the synthetic setup and
  \item microphone models learned from 15 seconds of paired audio from the re-recorded test set.
\end{enumerate*}
The last row of Table~\ref{table:speech-comm} shows that this model is significantly more robust on both modified test sets (94.5\% vs. 91.8\%, 91.0\% vs. 88.3\%).

\looseness -1 As baselines, we implemented spectral equalization~\cite{stockham1975blind} based on 15 seconds of unpaired audio and SpecAugment~\cite{park2019specaugment} with 2 masked  slices in both the time and frequency domain, each with up to 10 elements. While being oblivious to the distortion applied to the test samples, SpecAugment nevertheless closely matches the performance of spectral equalization by recovering $20\%$ and $26\%$ of the accuracy loss. We have also experimented with Mic2Mic~\cite{8732945}, but we were unable to tune the method to outperform the naive baseline of no augmentation. This is due to working with only 15 seconds of target audio, which proves insufficient for stable CycleGAN training.

For MicAugment, we reuse the models learned in the previous experiment for the synthetic setup, whereas for the real setup, we train MicAugment using a 15 second clip from the Librispeech corpus~\cite{panayotov2015librispeech} recaptured with the MacBook. By augmenting the training data with MicAugment,  up to $70\%$ of the accuracy loss can be recovered.

\section{Conclusion}
\looseness -1 We presented MicAugment, a novel approach for performing microphone style transfer given only a few seconds of target audio. The key component in the design of MicAugment is a novel non-linear microphone model, which relies on strong inductive biases and is easily identifiable. We evaluated the quality of the style transfer by means of downstream tasks, and the results show that our proposed method significantly outperforms existing baselines.

% References should be produced using the bibtex program from suitable
% BiBTeX files (here: strings, refs, manuals). The IEEEbib.bst bibliography
% style file from IEEE produces unsorted bibliography list.
% -------------------------------------------------------------------------
\bibliographystyle{abbrv}
\bibliography{bibliography}

\end{document}